\documentstyle[revtex]{aps}
\begin{document}
\draft
\begin{title}
Gamma rays and neutrinos from the Crab Nebula 
produced by pulsar accelerated nuclei
% \thanks{Preprint number ADP-AT-..... , submitted to {\it Phys. Rev. Lett.}
\end{title}
\author
%[W. Bednarek and R.J. Protheroe]
       {W. Bednarek$^*$ and R.J. Protheroe}
\begin{instit}
Department of Physics and Mathematical Physics, University
of Adelaide, Adelaide, SA 5005, Australia \\
$^*$Permanent address: University of \L\'od\'z, 90-236 \L\'od\'z, 
ul. Pomorska 149/153, Poland.  
\end{instit}

\begin{abstract}

We investigate the consequences of the acceleration of heavy
nuclei (e.g. iron nuclei) by the Crab pulsar.  Accelerated nuclei
can photodisintegrate in collisions with soft photons produced in
the pulsar's outer gap, injecting energetic neutrons which decay
either inside or outside the Crab Nebula.  The protons from
neutron decay inside the nebula are trapped by the Crab Nebula
magnetic field, and accumulate inside the nebula producing
gamma-rays and neutrinos in collisions with the matter in the
nebula. Neutrons decaying outside the Crab Nebula contribute to
the Galactic cosmic rays. We compute the expected fluxes of
gamma-rays and neutrinos, and find that our model could account
for the observed emission at high energies and may be tested by
searching for high energy neutrinos with future neutrino
telescopes currently in the design stage.
\end{abstract}


\narrowtext

The Crab Nebula is a well established $\gamma$-ray source with a
complex spectrum extending up to at least TeV
energies~\cite{no93}.  Its emission below a few tens of MeV is
interpreted as synchrotron emission by electrons with energies up
to $\sim 10^{15-16}$ eV in the Crab Nebula magnetic field of
strength $\sim 3\times 10^{-4}$ G~\cite{goul65}. These models
interpret the emission at higher energies as being due this same
population of electrons by inverse Compton scattering (ICS) of
synchrotron photons (SSC models).  The TeV $\gamma$-ray emission
might also originate near (but above) the Crab pulsar light
cylinder as a result of ICS scattering of infrared photons produced in
one of the pulsar's outer gaps by $e^{\pm}$ pairs escaping from
another outer gap~\cite{kchl91}.  The possible importance of
hadronic processes as sources of $\gamma$-rays inside the Crab
Nebula has also been considered by Cheng et al.~\cite{chetal90}
who calculated the high energy $\gamma$-ray fluxes assuming that
relativistic protons accelerated in the Crab pulsar outer gap
interact with the matter inside the Nebula, and note that this
process may contribute in the TeV $\gamma$-ray range.

In this Letter, we analyze the consequences of acceleration of
heavy nuclei in the pulsar magnetosphere as a possible mechanism
of energetic radiation from the Crab Nebula.  The importance of
photodisintegration of nuclei is determined by the reciprocal
mean free path which we calculate using cross sections of
Karaku\l a \& Tkaczyk~\cite{kt93} and show in Fig.~\ref{fig1} for
nuclei with atomic mass numbers between 4 and 56 during
propagation in the radiation field supposed to be present in the
outer gap of the Crab pulsar.  Ho~\cite{h89} has shown that the
total luminosity and spectrum of the outer gap radiation is not
strongly dependent on $P$.  For $P=33$~ms we adopt the photon
number density given by Eq.~7.1 in Paper~II of Cheng et
al.~\cite{chr86b}, and for other periods we scale this by $r_{\rm
LC}^{-3}$.  We assume this radiation is produced by cascading due
to electrons and positrons accelerated in the gap, and so is
highly anisotropic and directed across the gap (in both
directions).  The dimension of the outer gap is of the order of
the radius of the light cylinder, $l_{\rm gap}\approx r_{\rm LC}
= c/\Omega$.  For the case of the Crab pulsar, $r_{\rm LC}
\approx 1.5\times 10^{8}$ cm, and it is evident from
Fig.~\ref{fig1} that multiple photodisintegrations of primary Fe
nuclei will occur provided they can be accelerated to Lorentz
factors above $\sim 10^3$.  To check whether pion photoproduction
and $e^\pm$ pair production by Fe nuclei are important energy
loss mechanisms in the outer gap, we compute also the mean
interaction length for pion photoproduction using the cross
section given by Stecker~\cite{s68}, and the energy loss length
for pair production using the approximations given by Chodorowski
et al.~\cite{czs92}.  The results are shown in Fig.~\ref{fig1}.
It is clear that these processes are not important loss
mechanisms compared to photodisintegration, e.g the pair
production energy loss length is two orders of magnitude longer
than the length of the gap.

We assume that Fe nuclei can escape from the polar cap surface of
the Crab pulsar, and move along magnetic field lines to enter the
outer gap where they can be accelerated in the outer gap
potential as in the model of Cheng et al.~\cite{chr86b} and
Ho~\cite{h89}. The nuclei accelerated in the outer gap will
interact with photons either producing secondary $e^{\pm}$ pairs
(with negligible loss of energy) or extracting a nucleon.  The
pairs could eventually saturate the electric field of the gap
although this is far from certain.  We shall assume that for some
reason the field is not shorted.  Possibilities include: (a)
nuclei, having much larger Larmor radii and much lower
synchrotron and curvature losses than electrons, more easily
drift across the magnetic field lines away from the region of the
gap where they had produced pairs; (b) the secondary $e^{\pm}$
pairs may not be able to saturate the electric field in the outer
gap which is close to the light cylinder since the Goldreich \&
Julian~\cite{gj69} density tends to infinity where the gap
approaches the light cylinder.
 
In order to obtain the energy spectrum of neutrons extracted from
Fe nuclei, $N_n(\gamma_n)$, we simulate their acceleration and
propagation through the outer gap using a Monte Carlo method. To
obtain the rate of injection of neutrons per unit energy we
multiply $N_n(\gamma_n)$ by the number of Fe nuclei injected per
unit time, $\dot{N}_{\rm Fe}$, which can be simply related to the
total power output of the pulsar $L_{\rm Crab}(B,P)$~\cite{mt77},
\begin{eqnarray}
\dot{N}_{\rm Fe}= \xi L_{\rm Crab}(B,P)/Z \Phi(B,P),
\label{eq9}
\end{eqnarray}
\noindent
where $\xi$ is the parameter describing the fraction of the total
power taken by relativistic nuclei accelerated in the outer gap,
$Z=26$ is the atomic number of Fe,
$B$ is the surface magnetic field, $P$ is the pulsar's
period, and
\begin{eqnarray}
\Phi(B,P) \approx 5 \times 10^{16} \left( {B \over 4 \times 10^{12} \,
{\rm G}} \right) \left( {P \over 1 {\rm \, s}} \right)^{4/3} \hspace{1mm} \rm V
\label{eq:potential}
\end{eqnarray}
is the potential difference across the outer gap~\cite{chetal90}.
We assume $B = 4\times 10^{12}$ G.

Soon after the supernova explosion, when the nebula was
relatively small, nearly all energetic neutrons would be expected
to decay outside the nebula.  However, at early times we must
take account of collisions with matter.  The optical depth may be
estimated from $\tau_{nH}\approx \sigma_{pp} n_H r \approx
8.6\times 10^{14} M_1 v_8^{-2} t^{-2}$, where $M = M_1M_{\odot}$
is the mass ejected during the Crab supernova explosion in units
of solar masses, $r=vt$, $n_H = M/(4/3\pi r^3 m_p)$ is the number
density of target nuclei, and $v = 10^8 v_8$ cm s$^{-1}$ is the
expansion velocity of the nebula.  We note that $\tau_{nH} =
1$ at $t \approx 0.93 M_1^{1/2}v_8^{-1}$~y.

The number and spectrum of relativistic protons from neutron
decay at the present time is determinated by the injection rate
of neutrons into the Crab Nebula integrated over time since the
pulsar's birth. We estimate the evolution of the Crab pulsar's
period from birth to the present time taking account of magnetic
dipole radiation energy losses and gravitational energy losses
for an ellipticity of $3 \times 10^{-4}$~\cite{st83}. Magnetic
dipole losses determine the pulsar period at present, but the
initial period is determined largely by gravitational losses and
is probably shorter than $\sim 10$ ms.  Hence, we consider two
initial periods: $5$ ms, and $10$ ms.

The spectrum of protons from neutron decay outside the Crab
Nebula is given by
\begin{eqnarray}
N_p^{\rm out} (\gamma_p, t_{\rm CN}) &=& \nonumber \\ 
& & \hspace{-2.1truecm} \int_0^{t_{\rm CN}} {\rm d}t
\dot{N}_{\rm Fe}(t) N_n(\gamma_p,t) e^{-\tau_{nH}(t)}
e^{-vt_{\rm CN}/c\gamma_p\tau_n}
\label{eq10}
\end{eqnarray}
where $\tau_n\approx 900$ seconds is the neutron decay time, and we
make the approximation that the Lorentz factor of protons is
equal to that of parent neutrons, $\gamma_p \approx \gamma_n$.

The equation for the spectrum of protons injected inside the Crab
Nebula is more complicated because we must take account of
proton-proton collisions and adiabatic energy losses due to the
expansion of the nebula.  The Lorentz factor of these protons at
time $t$ after the explosion such that their present Lorentz factor
is $\gamma_p$ is given by
\begin{eqnarray}
\gamma_p(t) \approx 
\gamma_p {(t+t_{\rm CN}) \over 2tK^{\tau_{pp}(t) }}
\label{eq:Lorentz_factor}
\end{eqnarray}
where $\tau_{pp}(t)$ is the optical depth for collision of
protons with matter between $t$ and $t_{\rm CN}$, and is given
by $\tau_{pp}(t)\approx 1.3\times 10^{17} M_1 v_8^{-3} (t^{-2} -
t^{-2}_{\rm CN})$, and $K\approx 0.5$ is the inelasticity
coefficient in proton--proton collisions.

Since we are interested in protons interacting inside the nebula at
the present time, we must also include those neutrons which
decayed at locations outside the nebula at time $t$ but which
will be inside the nebula at time $t_{\rm CN}$.  Taking account of
all these effects, we arrive at the formula below for the proton
spectrum inside the nebula at time $t_{\rm CN}$, 
\begin{eqnarray}
\hspace{-0.7truecm}
N_p^{\rm in} (\gamma_p, t_{\rm CN}) & = &  \gamma_p^{-1} \int_0^{t_{\rm CN}} 
 {\rm d}t \dot{N}_{\rm Fe}(t) 
e^{-\tau_{nH}(t)}   \nonumber \\
& &  
\hspace{-2.1truecm}
\times \biggl[ N_n(\gamma_p(t),t) \gamma_p(t)
[ 1 - \exp({-vt/c\gamma_p(t)\tau_n})]    \nonumber \\
& & 
\hspace{-2.1truecm}
+  
\int_{t}^{t_{\rm CN}} {\rm d}t' N_n(\gamma_p(t'),t)
\gamma_p(t')
{{v \exp({-vt'/c\gamma_p(t')\tau_n}})\over{c\gamma_p(t') \tau_n}} 
\biggr].  
\label{eq10a}
\end{eqnarray}
The first term gives the contribution from neutrons decaying
initially inside the nebula while second term gives the
contribution from neutrons decaying at points initially outside
the nebula which will be inside the nebula at time $t_{\rm CN}$.

We compute separately the spectra of protons from neutrons
decaying inside and outside the Crab Nebula.  In Fig.~\ref{fig3}
proton spectra are shown for two initial periods and present
nebula radii of $1$ pc and $2$ pc.  To check if protons decaying
inside will remain near where they were produced, or diffuse into
the Galactic environment, we estimate their typical diffusion
distance during time $t_{\rm CN}$, $x_{\rm dif}\approx (c r_L
t_{\rm CN}/3)^{1/2}$, where $r_L = m_p \gamma_p c^2/e B$ is the
Larmor radius, and we have used the minimum diffusion
coefficient.  For protons at the peak of the energy spectrum with
$\gamma_p = 10^{5}$ in a magnetic field of the order of $B =
5\times 10^{-6}$ G, the typical diffusion distance is $x_{\rm
dif}\approx 1$ pc, which is comparable to the radius of the Crab
Nebula.  Hence protons which decayed inside, were probably
trapped and accumulated inside the nebula.  Protons from neutrons
decaying outside the nebula will typically have higher energies
and diffuse farther, thus escaping to become Galactic cosmic
rays.

The protons which have accumulated inside the Crab Nebula since
the pulsar was formed can produce observable fluxes of
$\gamma$-rays and neutrinos.  We compute the expected
$\gamma$-ray spectra for four different models, taking various
initial pulsar periods, present sizes and masses of the Crab
Nebula. The fluxes may possibly be enhanced if protons are
efficiently trapped by the dense filaments as suggested by Atoyan
and Aharonian \cite{Atoyan96}.  Filaments with density $\sim 500$
cm$^{-3}$ are present in the Crab Nebula~\cite{df85}. Therefore
we introduce an effective density experienced by the protons
inside nebula, $n_H^{\rm eff} = \mu n_H$, where $n_H$ is defined
above, and the parameter $\mu$ takes into account the possible
effects of proton trapping by the filaments.  The $\gamma$-ray
spectra are computed in terms of a scaling model~\cite{hil81},
and the photon fluxes expected at the Earth are shown in
Fig.~\ref{fig5} for $\xi \mu = 1$, and are compared with
observations of the Crab Nebula above 0.2 TeV.  Results are shown
for models I to IV having $P_0$, $r_{\rm CN}$ and $M_1$ as
specified in Table~\ref{tab1}, and for a distance to the Crab Nebula of
1830 pc~\cite{df85}.  Comparison with the Whipple observations at
10 TeV allows us to place constraints on the free parameters of
the model, and upper limits on $\xi\mu$ which are given in
Table~\ref{tab1} as $(\xi \mu)_\gamma$. Note that it is usually
argued for the standard pulsar model that the rate of injection
of Fe nuclei into the pulsar magnetosphere should not cause the
charge density to exceed the Goldreich \& Julian
density~\cite{gj69}.  In the case of no additional currents
flowing through the magnetosphere, this condition constrains the
value of $\xi$ to $\xi \approx 0.1$ for the Crab pulsar with
period 33~ms.  However, in the presence of additional currents
the value of $\xi$ can be higher.

The question of the importance of hadronic interactions in the
Crab Nebula can be settled by the detection of a neutrino signal
from the Crab.  In Fig.~\ref{fig6} we show the neutrino spectrum
produced in collisions of protons with matter inside the Crab
nebula for the models considered above, and compare this with the
atmospheric neutrino background flux within $1^\circ$ of the
source direction~\cite{lip93}.  It is clear that neutrino
detectors with good angular resolution should be able to detect
neutrinos at 10 TeV from the Crab Nebula if $\xi \mu$ is greater than 
$(\xi \mu)_\nu$ given in Table~\ref{tab1}. Note that in all cases 
$(\xi\mu)_\nu < (\xi\mu)_\gamma$ and so the possibility of $\nu$
detection is allowed by the existing $\gamma$-ray observations.

In the present paper we have not considered those nuclei which had
survived propagation through the outer gap radiation field and
were injected into the inner nebula region following magnetic
field lines. These nuclei (and protons) are expected to 
accumulate in the inner part of the Crab Nebula were the
magnetic fields are $\sim 2\times 10^{-3}$~G 
measured in the synchrotron wisps which are located at $\sim 0.2$
pc~\cite{he95}.  The diffusion distance scale during the age of
the Crab Nebula is $\sim 0.3$~pc
for a Lorentz factor $\gamma_A = 10^6$ and $B = 10^{-3}$ G. This
means that most nuclei will probably be confined to the inner part of the
Crab Nebula where there is no evidence of dense
matter~\cite{df85} (other than the neutron star).  Hence, we may
be justified in neglecting the contribution of such nuclei to the
$\gamma$-ray and neutrino production by hadronic collisions in
comparison with the contribution from interactions of protons from
neutron decay.  However, some fraction of the energy of these
nuclei could be transferred to very high energy electrons by
resonant scattering~\cite{hoetal92}.  Such electrons might then
produce additional $\gamma$-rays by the SSC process as in the models
discussed in ref.~\cite{goul65}. Hence, we suggest that the complex Crab
Nebula spectrum may in fact be formed as a result of both
processes, i.e. SSC and as a result of pulsar acceleration of
nuclei discussed here.

Another consequence of our model is that particles (nuclei and
protons from neutron decay) will be also injected by the Crab
Pulsar and other pulsars into the Galactic cosmic rays, this may
effect the cosmic ray spectrum and composition.  This problem
needs closer investigation, and should take into account the
parameters of the Galactic population of pulsars at birth.
However, it is worth noting that the energy distribution of the
protons from decay of neutrons outside the Crab nebula peaks at
$\sim 10^{15}$ eV (see Fig.~\ref{fig3}), which is close to the
energy of the knee in the cosmic ray spectrum.

W.B. thanks the Department of Physics and Mathematical Physics at the 
University of Adelaide for hospitality during his visit. 
This research is supported by a grant from the Australian Research Council.

\begin{table}
\caption{Model parameters and limits on $\xi\mu$.}
\begin{tabular}{lrrrr}
Model&I&II&III&IV\\
\tableline
$P_0$ (ms) & 5 & 10 & 10 & 10 \\
$r_{\rm CN}$ (pc) & 1 & 2 & 2 & 1 \\
$M_1$ & 3 & 3 & 10 & 3 \\
$(\xi \mu)_\gamma$ at 10 TeV & 0.63 & 6.9 & 2.0 & 1.0 \\
$(\xi\mu)_\nu$ at 10 TeV & 0.22 & 1.8 &  0.54 & 0.29 \\
\end{tabular}
\label{tab1}
\end{table}

\newpage
~\\

\vspace{7.3cm} 
\figure{ 
\includegraphics{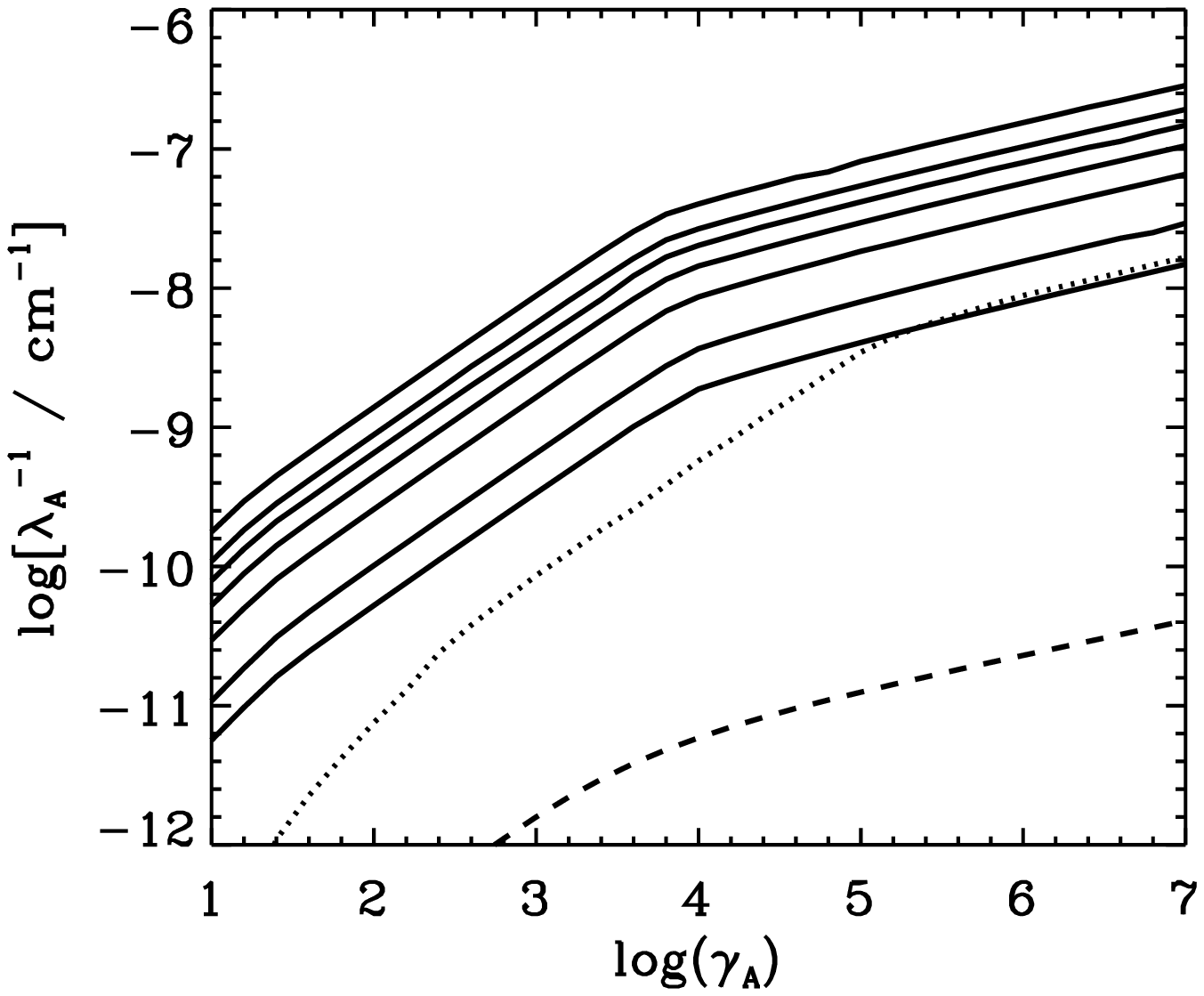}
Reciprocal mean free paths for photodisintegration of nuclei with
mass numbers, A = 56, 40, 32, 24, 16, 8, and 4 (full curves from
the top) in the radiation field of the Crab pulsar's outer
gap as a function of the Lorentz factor of the
nuclei.  The reciprocal mean free path for pion photoproduction
(dotted curve) and the energy loss length for $e^\pm$ pair
production (dashed curve) by iron nuclei are also shown.
\label{fig1}}

\vspace{7.3cm} 
\figure{ 
\includegraphics{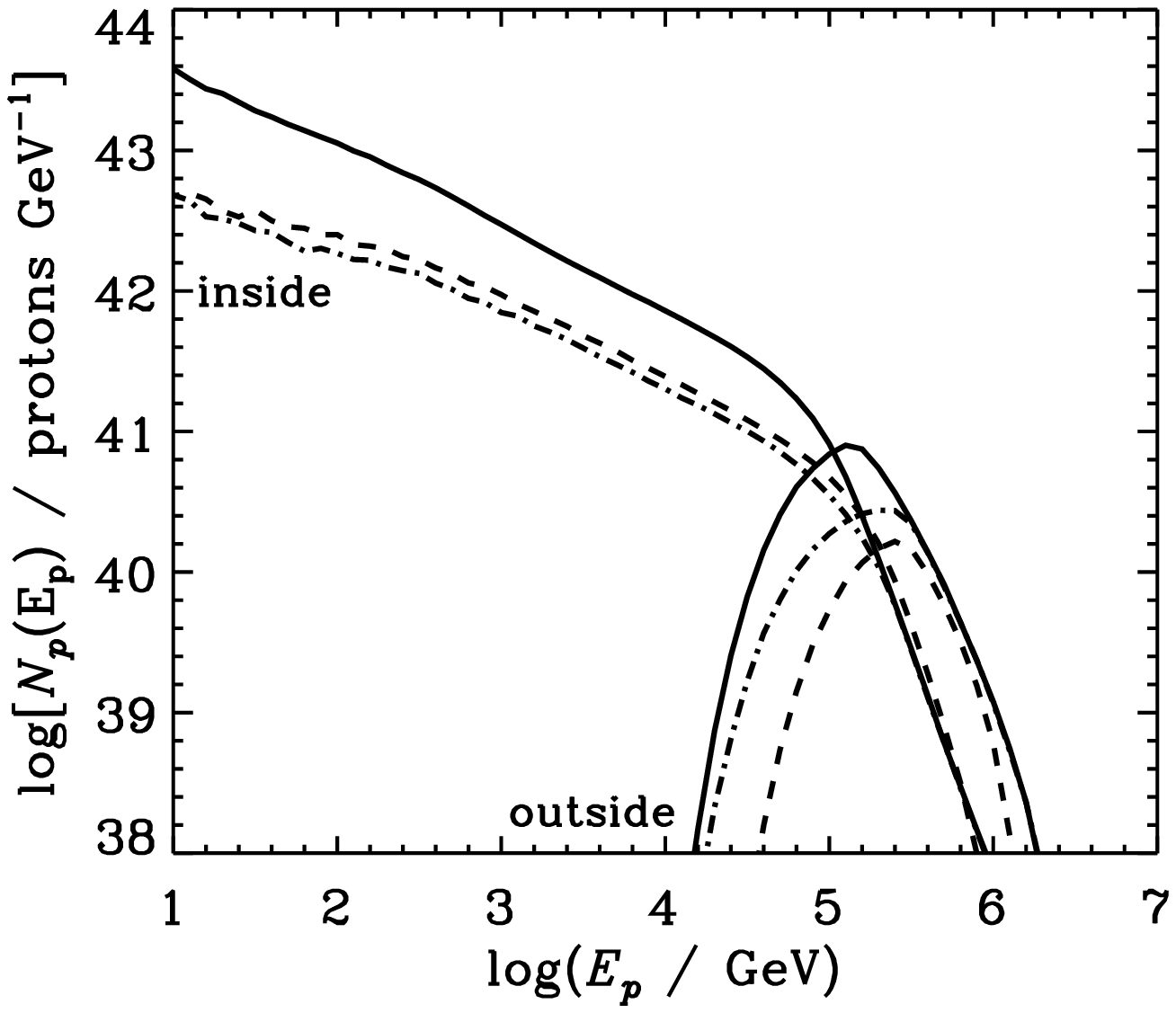} 
Spectra of protons from the decay of neutrons injected by the
Crab pulsar inside and outside the Crab Nebula assuming $M_1=3$
for the following parameters: $P_0 =10$ ms, and Crab Nebula
radius $r_{\rm CN}=1$ pc (dot-dashed curve) and 2 pc (dashed
curve), and for $P_0 = 5$ ms and $r_{\rm CN} = 1$ pc (full
curve).
\label{fig3}}

\newpage
 ~\\
\vspace{7.3cm} 
\figure{ 
\includegraphics{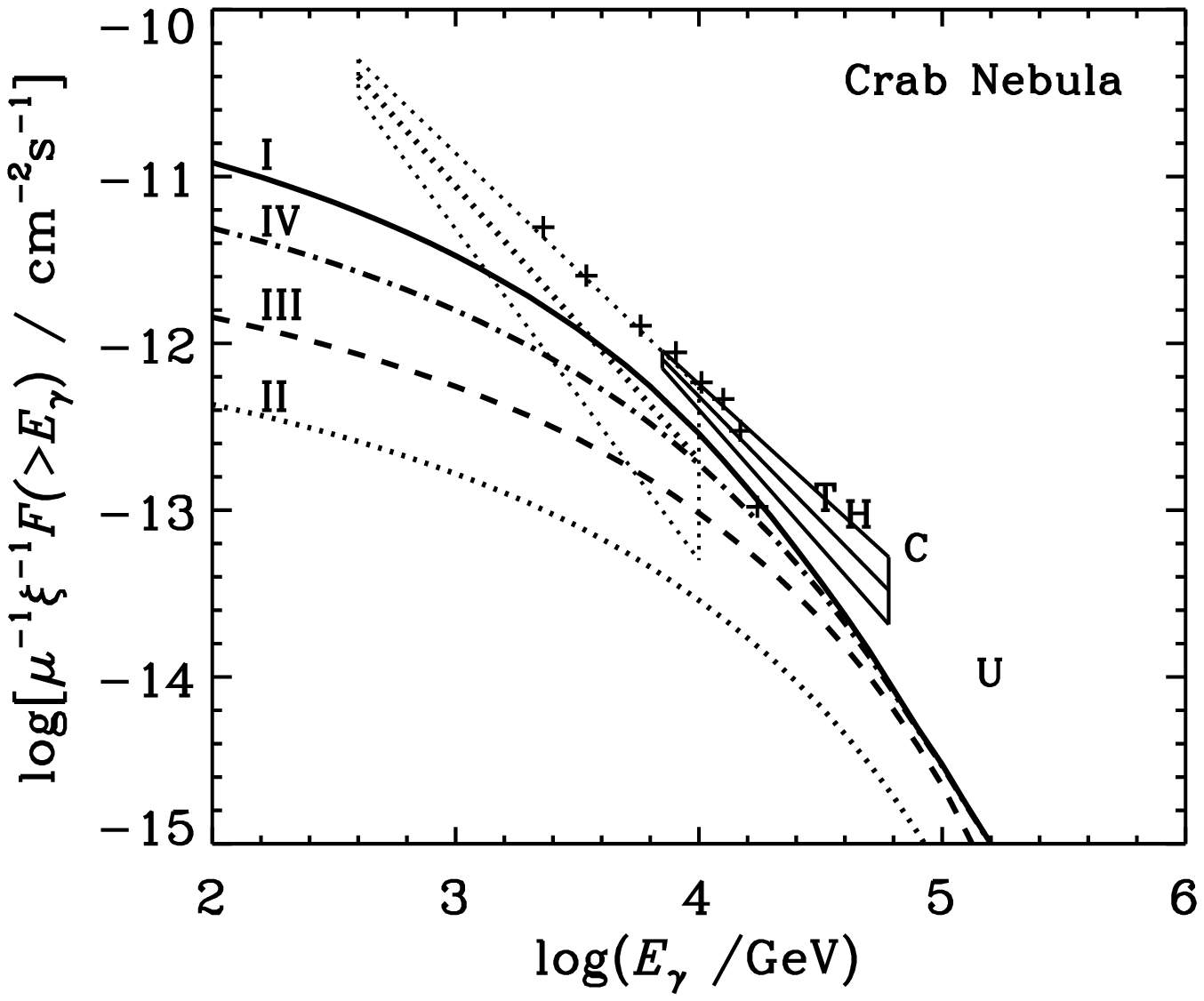} 
Spectra of $\gamma$-rays from interactions of protons with matter
inside the Crab Nebula for the different proton spectra shown in
Fig~\ref{fig3} and for the different models of the nebula (I --
IV) considered in the text.  Observations: Whipple
Observatory~\cite{letal93} (dotted line and error box);
THEMISTOCLE~\cite{themis95} (+); and CANGAROO~\cite{Tanimori97}
(solid line and error box).  Upper limits from various
experiments mentioned in ref.~\cite{ameetal95}: T - Tibet, H - HEGRA,
C - CYGNUS, and U - CASA-MIA.
\label{fig5}}

\vspace{7.3cm} 
\figure{ 
\includegraphics{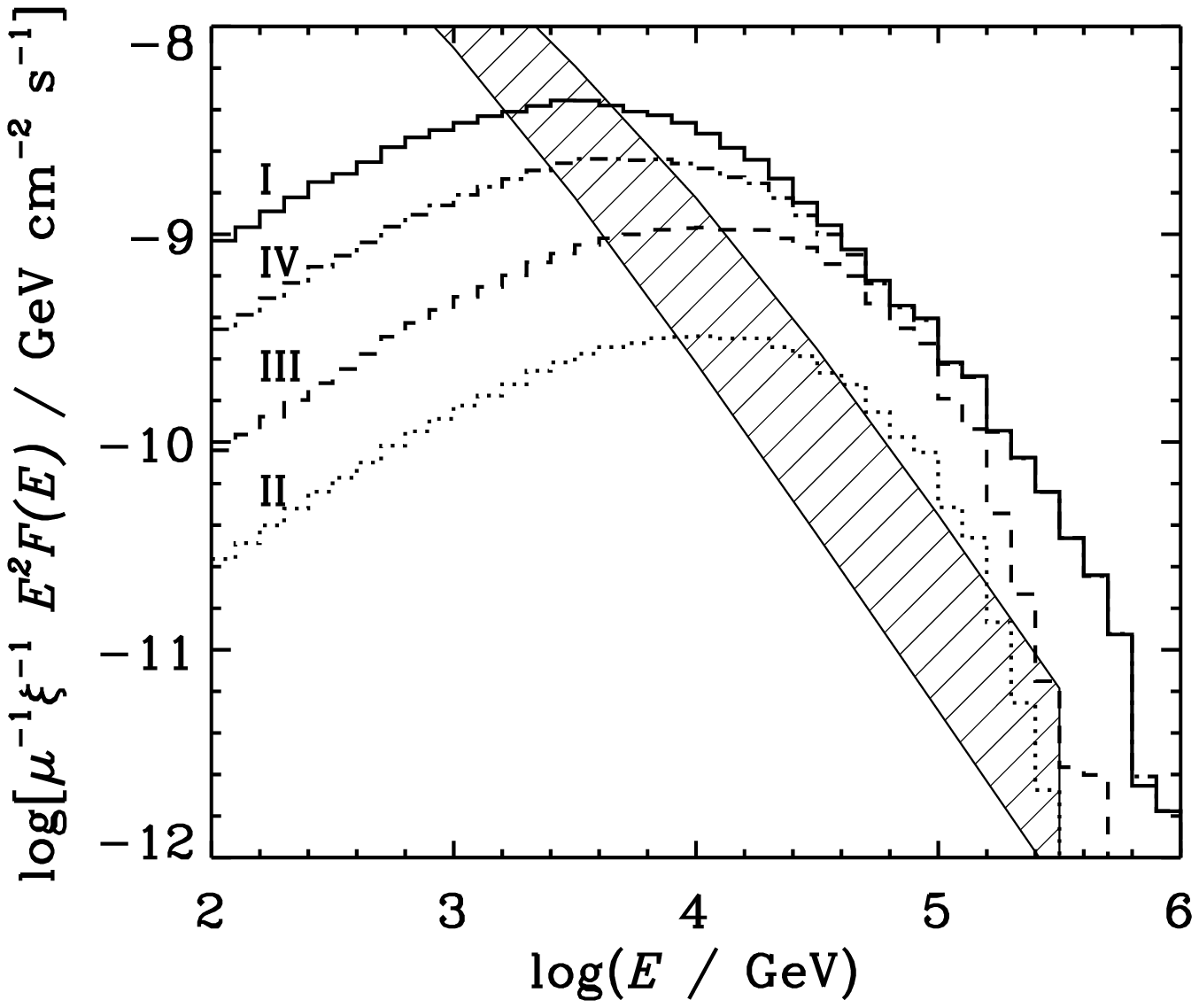}
 Spectra of neutrinos ($\nu_\mu + \bar{\nu}_\mu$) expected for the
different models (I - IV) considered in the text. The atmospheric
neutrino background~\cite{lip93} within $1^\circ$ of the source
is indicated by the hatched band which shows the variation with
angle: horizontal (upper bound), vertical (lower bound).
\label{fig6}}

\end{document}